\documentclass[journal]{IEEEtran}
\usepackage{amsmath, amsfonts, amsthm, bm}
\usepackage[T1]{fontenc}
\usepackage{algorithmic}
\usepackage{algorithm}
\usepackage{array}
\usepackage{textcomp}
\usepackage{stfloats}
\usepackage{url}
\usepackage{verbatim}
\usepackage{graphicx}
\usepackage{cite}
\usepackage{epsfig}
\usepackage{color,soul}
\usepackage{multicol}
\usepackage{multirow}
\usepackage{amssymb}
\usepackage{hhline}
\usepackage{subcaption}

\begin{document}

\title{Highly Efficient Parallel Row-Layered Min-Sum MDPC Decoder for McEliece Cryptosystem}
\author{Jiaxuan Cai and Xinmiao Zhang
\thanks{This material is based upon work supported by the National Science Foundation under Award No. 2052641. The authors are with the Department of Electrical and Computer Engineering, The Ohio State University, Columbus, OH 43210 USA (E-mails: cai.1072@osu.edu; zhang.8952@osu.edu).}}

\markboth{CAI AND ZHANG: Highly Efficient Parallel Row-Layered Min-Sum MDPC Decoder for McEliece Cryptosystem}%
{}
\maketitle

\begin{abstract}
The medium-density parity-check (MDPC) code-based McEliece cryptosystem remains a finalist of the post-quantum cryptography standard. The Min-sum decoding algorithm achieves better performance-complexity tradeoff than other algorithms for MDPC codes. However, the prior Min-sum MDPC decoder requires large memories, whose complexity dominates the overall complexity. Besides, its actual achievable parallelism is limited. This paper has four contributions: For the first time, the row-layered scheduling scheme is exploited to substantially reduce the memory requirement of MDPC decoders; A low-complexity scheme is developed to mitigate the performance loss caused by finite precision representation of the messages and high column weights of MDPC codes in row-layered decoding; Constraints are added to the parity check matrix construction to enable effective parallel processing with negligible impacts on the decoder performance and resilience towards attacks; A novel parity check matrix division scheme for highly efficient parallel processing is proposed and the corresponding parallel row-layered decoder architecture is designed. The number of clock cycles for each decoding iteration is reduced by a factor of $L$ using the proposed $L$-parallel decoder with very small memory overhead. For an example 2-parallel decoder, the proposed design leads to 26\% less memory requirement and 70\% latency reduction compared to the prior decoder.
\end{abstract}


\section{Introduction}

The National Institute of Standards and Technology (NIST) has initiated the standardization of post-quantum cryptography in response to the imminent needs for cryptographic schemes that can withstand quantum computing attacks. It announced the fourth round of finalists in July 2022. The McEliece cryptosystem that employs quasi-cyclic medium-density parity-check (QC-MDPC) codes \cite{McElieceNIST,MDPCMcEliece} remains one of the candidates. The design and implementation of low-density parity-check (LDPC) decoders for error correction in storage systems and digital communication have been extensively explored. The parity check matrices of popular QC-LDPC codes consist of a large number of smaller cyclic permutation matrices (CPMs). Efficient parallel decoding can be achieved by processing one or multiple CPMs at a time \cite{ParaLayeredLDPCArch1,ParaLayeredLDPCArch2,ParaLayeredLDPCArch3}. However, the parity check matrices for the MDPC codes considered for the McEliece cryptography scheme consist of a few very large circulant matrices, whose nonzero entries are randomly generated and their column weights are much higher. As a result of the irregularity and high column weight, many prior techniques for simplifying LDPC decoders do not apply to MDPC decoders. The decoding algorithms require re-evaluation and the decoder architectures need to be re-designed.

A spectrum of algorithms can be used to decode MDPC codes, with the bit-flipping (BF) algorithm and its variations being the simplest and having been explored in quite a few recent works \cite{Liva, BIKE, Baldi,ErrorFloor}. The BF MDPC decoder implementations in \cite{GuneysuDATE, Guneysu, Hu} divide each column of the parity check matrix into $L$-bit segments, which are processed one by one to reduce the decoding latency. However, the parity check matrices of MDPC codes are still very sparse. Taking this into account, the decoder in \cite{XieSparse} only processes the nonzero segments and utilizes an out-of-order processing scheme to reduce the complexity, decoding latency, and memory access.

MDPC decoding algorithms with lower failure rates can better resist reaction-based attacks that try to recover the secret parity check matrix of the code by utilizing decoding failures \cite{BaldiAttack}. One of the most powerful attacks is the Guo-Johansson-Stankovski (GJS) attack \cite{Reaction}. The Min-sum algorithm \cite{Minsum} with optimized scalars \cite{McElieceSlicedMP} achieves orders of magnitude lower decoding failure rate compared to BF algorithms. The Min-sum MDPC decoder proposed in \cite{McElieceSlicedMP} divides the parity check matrix into $L$ blocks of rows and processes one nonzero entry in each block row in parallel to increase the efficiency and throughput. Because of the high column weight and long length of the MDPC codes, the decoder has large memory requirement and the memories dominate the overall decoder complexity. Besides, due to the uneven distribution of the nonzero entries in the randomly constructed parity check matrix, the speedup achieved by the decoder in \cite{McElieceSlicedMP} is far less than $L$ and does not improve much for larger $L$ despite the much bigger memory overhead.

Unlike the sliced message-passing scheduling scheme \cite{SlicedLDPC} used in the design of \cite{McElieceSlicedMP} that updates the messages at the end of each decoding iteration, the row-layered scheduling scheme \cite{LayeredLDPC} utilizes the most updated messages in the remainder of the decoding iteration. As a result, both the number of decoding iterations and memory requirement are reduced. Row-layered decoding has been extensively studied for LDPC codes \cite{ParaLayeredLDPCArch1,ParaLayeredLDPCArch2,ParaLayeredLDPCArch3}. However, the high column weights of MDPC codes lead to decoding failure rate increase when finite precision is adopted in the hardware implementation of the decoder. Additionally, the random structure of their parity check matrices makes it difficult to achieve efficient parallel row-layered decoding.

For the first time, this paper investigates the design of efficient parallel row-layered Min-sum decoders for the MDPC codes considered in the McEliece cryptosystem. The major contributions of this paper include the following: 
\begin{itemize}
    \item A serial row-layered decoder is first proposed for MDPC codes as the baseline for our parallel decoder architecture. The decoder processes only the nonzero entries of the parity check matrix and only the locations of the nonzero entries are stored to reduce the memory requirement and latency. By processing the parity check matrix layer by layer, the size of the memory storing the check-to-variable (c2v) messages and accordingly the overall memory requirement are substantially reduced. 
    \item The decoding failure rate increase caused by finite precision representation of the scaled c2v messages is analyzed. The performance gap is eliminated by keeping fractional bits in the messages with small  overheads on memory requirement and critical path.
    \item Constraints are proposed in the random construction of the parity check matrices for MDPC codes to enable effective parallel processing. Analyses and simulations have been carried out on the possible effects of the constraints on the number of possible keys, decoding failure rate, and resilience towards reaction attacks.
    \item A novel dynamic parity check matrix division scheme based on the locations of identity blocks is proposed to achieve highly efficient parallel processing. The corresponding parallel row-layered decoder architecture is also designed.
\end{itemize}

The proposed $L$-parallel decoder effectively reduces the number of clock cycles needed for each decoding iteration by a factor of $L$. Additionally, the memory overhead associated with parallel processing is very small. As a result, the proposed design can efficiently exploit a much higher parallelism such as $32$ instead of only $2$ or $4$ as in prior design \cite{McElieceSlicedMP}. For an example MDPC code considered for the standard, the proposed $2$-parallel row-layered Min-sum decoder reduces the size of the memories, which dominate the decoder complexity, by 26\% and shorten the latency by 70\% compared to that in \cite{McElieceSlicedMP}. 

This paper is organized as follows. Section II introduces background information. Section III proposes the overall architecture of a serial row-layered Min-sum decoder and methodologies to eliminate the decoding performance gap caused by finite precision representation of messages. In Section IV, constraints for MDPC code construction for enabling efficient parallel processing is presented and it is shown that they do not have significant effects on decoding performance and possible attacks. The dynamic parity check matrix division scheme for highly efficient parallel processing and the corresponding implementation architecture are detailed in Section V. Section VI presents the hardware complexity analyses and comparisons, and conclusions follow in Section VII.

\section{Background}

\subsection{MDPC codes and McEliece cryptosystem}
An MDPC code is a type of linear block code, which can be defined using a parity check matrix denoted as $\mathbf H$. A vector $\mathbf c$ is a valid codeword if and only if $\mathbf c\mathbf H^T=0$. The private key of the McEliece cryptosystem based on MDPC codes is a set of $n_0$ $r$-bit random vectors, each with a Hamming weight of $w$. The parity check matrix of the MDPC code is in the format of $\mathbf H=[\mathbf H_0|\mathbf H_1|\cdots|\mathbf H_{n_0-1}]$, where $\mathbf H_i$ ($0\leq i<n_0$) is a circulant matrix with the first column equaling the $i$-th random vector. If $\mathbf H_{n_0-1}$ is non-invertible, the corresponding vector is randomly regenerated. The generator matrix $\mathbf G$ and the parity check matrix $\mathbf H$ of a linear block code satisfy $\mathbf G \mathbf H^T=0$. Consequently, the $\mathbf G$ matrix for the McEliece cryptosystem is derived as $[\mathbf I|\mathbf B^T]$, where $\mathbf B=[\mathbf H_{n_0-1}^{-1}\mathbf H_0|\mathbf H_{n_0-1}^{-1}\mathbf H_1|\cdots|\mathbf H_{n_0-1}^{-1}\mathbf H_{n_0-2}]$. The first columns of circulant matrices $\mathbf H_{n_0-1}^{-1}\mathbf H_i$ ($0\leq i< n_0-1$) constitute the public key in the McEliece cryptosystem.

\begin{figure}[t] 
    \begin{center}
    \includegraphics[width=.45\textwidth]{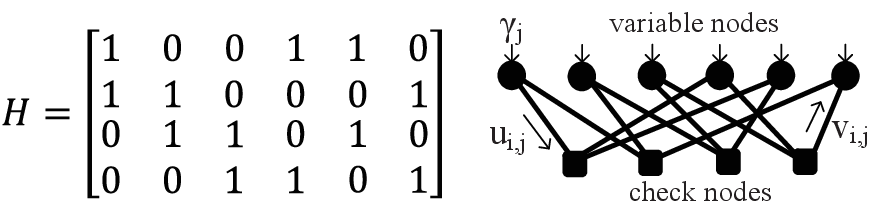}
    \vspace {-0em}\caption{ The $\mathbf H$ matrix and the corresponding Tanner graph of a toy MDPC code.}\label{Tanner}
    \end{center}
    \vspace {-0.5em}
\end{figure}

A bipartite graph called Tanner graph can be used to represent the $\mathbf H$ matrix of an MDPC code. This graph has two types of nodes known as variable nodes and check nodes. The $\mathbf H$ matrix and the corresponding Tanner graph of a toy MDPC code is depicted in Fig. \ref{Tanner}. Each variable node corresponds to a column of $\mathbf H$, while each check node represents a row. A pair of variable and check nodes are connected by an edge if the corresponding entry in the $\mathbf H$ matrix is nonzero.

\begin{figure}[t] 
    \begin{center}
    \includegraphics[width=.45\textwidth]{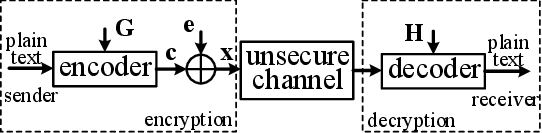}
    \vspace {-0em}\caption{ The encryption and decryption process of the McEliece cryptosystem based on QC-MDPC codes.}\label{endecrypt}
    \end{center}
    \vspace {-0.5em}
\end{figure}

\begin{table}[t]
    \renewcommand\arraystretch{1.15}
    \caption{QC-MDPC code parameters considered for the post-quantum cryptography standard \cite{MDPCMcEliece}.}\label{para} \vspace{-1em}
    \begin{center}
    \begin{tabular}{c|c|c|c|c|c}
    \hline
    security level (bits)& $n_0$ & $n$ &$r$ & $w$ & $t$ \\\hline\hline
    80 & 2 & 9602 & 4801 & 45 & 84\\
    & 3 & 10779 & 3593 & 51 & 53\\
    & 4 &12316 & 3079 & 55 & 42 \\\hline
    128 & 2 & 19714 & 9857 & 71 & 134\\
    & 3 & 22299 & 7433 & 81 & 85\\
    & 4 &27212 & 6803 & 85 & 68 \\\hline
    256 & 2 & 65542 & 32771 & 137 & 264\\
    & 3 & 67593 & 22531 & 155 & 167\\
    & 4 &81932 & 20483 & 161 & 137 \\\hline
    \end{tabular}
    \end{center}
\end{table}

As shown in Fig. \ref{endecrypt}, the core process of encryption in the McEliece cryptosystem is MDPC encoding. An $(n_0-1)r$-bit plaintext vector is multiplied with the generator matrix $\mathbf G$ to compute a codeword of $n=n_0r$ bits denoted as $\mathbf c$. Then the ciphertext $\mathbf x$ is derived by adding a randomly generated $n$-bit vector $\mathbf e$ containing $t$ nonzero bits to $\mathbf c$. The decryption is to perform MDPC decoding on the ciphertext $\mathbf x$. If $\mathbf x$ is decodable, then $\mathbf c$ is recovered. Since the $\mathbf G$ matrix is systematic, the first $(n_0-1)r$ bits of the codeword $\mathbf c$ equal the plaintext. The parameters of the MDPC codes considered for the post-quantum cryptography standard are listed in Table \ref{para}.

\subsection{Min-sum decoding algorithm}

Algorithm \ref{minsumalgo} lists the pseudo codes for the Min-sum algorithm. It iteratively exchanges reliability information of multiple bits between connected variable and check nodes to make decisions on the received bits until a codeword is found. For an input vector $\mathbf x=[x_0,x_1,\cdots,x_{n-1}]$, the initial reliability information for $x_j$, denoted as $\gamma_j$, is configured as either $+C$ or $-C$, depending on whether $x_j$ is `0' or `1', respectively. Simulations can be used to determine the optimal value of $C$, which depends on the maximum value of the reliability messages and the column weight of the code. In the remaining part of this paper, $u_{i,j}$ denotes the variable-to-check (v2c) message from variable node $j$ to check node $i$, and $v_{i,j}$ represents the c2v message from check node $i$ to variable node $j$. $k$ is the decoding iteration number in Algorithm \ref{minsumalgo}. 
$S_v(i)$ ($S_c(j)$) represents the set of variable (check) nodes connected to the check (variable) node $i$ ($j$). A sign bit of `0' and `1' indicates that the message is positive and negative, respectively, and $\oplus$ denotes the XOR operation. To reduce the decoding failure rate, the sum of c2v messages is multiplied with a scalar, $\alpha$, for variable node processing and computing {\it a posteriori} information, $\tilde{\gamma_j}$ \cite{Minsum}. If no valid codeword is found after $I_{\text{max}}$ iterations, a decoding failure is declared.

\begin{algorithm}
    \caption{Scaled Min-sum Decoding Algorithm \cite{McElieceSlicedMP}}
    \label{minsumalgo}
    \begin{algorithmic}[1]
\STATE {\bf Input:} $\mathbf x=[x_0,x_1,\cdots, x_{n-1}]$
    \STATE {\bf Initialization:} $u_{i,j} = \gamma_j$, $flag$= FALSE
    \STATE{\bf if}  $\mathbf x\mathbf H^T = 0$, \textbf{then}
\STATE$\quad$ $\quad$ $flag$=TRUE
\STATE$\quad$ $\quad$ goto \textbf{output}

    \FOR{$k = 1$ to $I_{\text{max}}$}
    
    \STATE $\quad$ {\bf Check node processing:}
    \STATE $\quad$ $min1_i = \min_{j\in S_v(i)} |u_{i,j}|$
    \STATE $\quad$ $idx_i = \text{arg} \min_{j\in S_v(i)} |u_{i,j}|$
    \STATE $\quad$ $min2_i = \min_{j\in S_v(i), j\neq idx_i} |u_{i,j}|$
    \STATE $\quad$ $s_i = \oplus_{j\in S_v(i)}\text{sign}(u_{i,j})$
    \STATE $\quad$   $\text{for each} \quad j \in S_v(i)$
    \STATE $\quad$ $\quad$$\quad$ $|v_{i,j}| = \begin{cases}
    min1_i &\text{if } j \neq idx_i \\
    min2_i &\text{if } j = idx_i 
    \end{cases}$
    \STATE $\quad$ $\quad$$\quad$ $sign(v_{i,j})=s_i\oplus sign(u_{i,j})$
    \STATE $\quad$ {\bf Variable node processing:}
    \STATE $\quad$ $u_{i,j} = \gamma_j + \alpha \sum_{i' \in S_c(j), i'\neq i} v_{i',j}$
    \STATE $\quad$ {\bf A posteriori info. comp. \& tentative decision:}
    \STATE $\quad$ $\tilde{\gamma_j} = \gamma_j + \alpha \sum_{i \in S_c(j)} v_{i,j}$
    \STATE $\quad$ $x_j = \text{sign}(\tilde \gamma_j)$
    \STATE $\quad$ \textbf{if} $\mathbf x\mathbf H^T = 0$, \textbf{then}
    \STATE $\quad$ $\quad$ $flag$ = TRUE
    \STATE $\quad$ $\quad$ goto \textbf{output}
    \ENDFOR
    \STATE \textbf{output:} $\mathbf x$
\STATE\textbf{if} $flag=$ FALSE, \textbf{then} declare decoding failure

    \end{algorithmic}
\end{algorithm}

\subsection{Row-layered scheduling scheme}

In row-layered decoding \cite{LayeredLDPC}, the parity check matrix $\mathbf H$ is divided into blocks of $f$ rows. Each block of $f$ rows is called a layer, and each column in a block should only have at most one nonzero entry. During the decoding of layer $l$ of iteration $k$, for a variable node, denote the corresponding v2c and c2v messages by $u^{(k,l)}$ and $v^{(k,l)}$, respectively. The notations omit the indices for both variable and check nodes for brevity. In row-layered decoding, the c2v messages computed from a layer are used right away to update the {\it a posterior} information and v2c messages used in the decoding of the next layer. Since all the c2v messages are initially zero, the initial {\it a posterior} information equals the channel information $\gamma$. Then after the decoding of layer $l$ in iteration $k$, the {\it a posterior} information is updated as
\begin{equation}\label{apost}
    \tilde{\gamma}' = \tilde{\gamma} - \alpha v^{(k-1,l)} + \alpha v^{(k,l)}.
\end{equation}
Comparing Lines 14 and 16 of Algorithm \ref{minsumalgo}. The v2c message can be calculated as
\begin{equation}\label{v2c}
u^{(k,l)} = \tilde \gamma - \alpha v^{(k-1,l)}.
\end{equation}

The row-layered Min-sum decoding algorithm is the same as that listed in Algorithm \ref{minsumalgo}, except that Lines 16 and 18 are replaced by \eqref{v2c} and \eqref{apost}, respectively. Since updated messages are utilized in the remainder of the same decoding iteration instead of the next iteration, the row-layered decoding converges faster. As a result, it also achieves better decoding performance  compared to Algorithm \ref{minsumalgo} when $I_{\text {max}}$ is limited.

\begin{figure}[t] 
    \begin{center}
    \includegraphics[width=.47\textwidth]{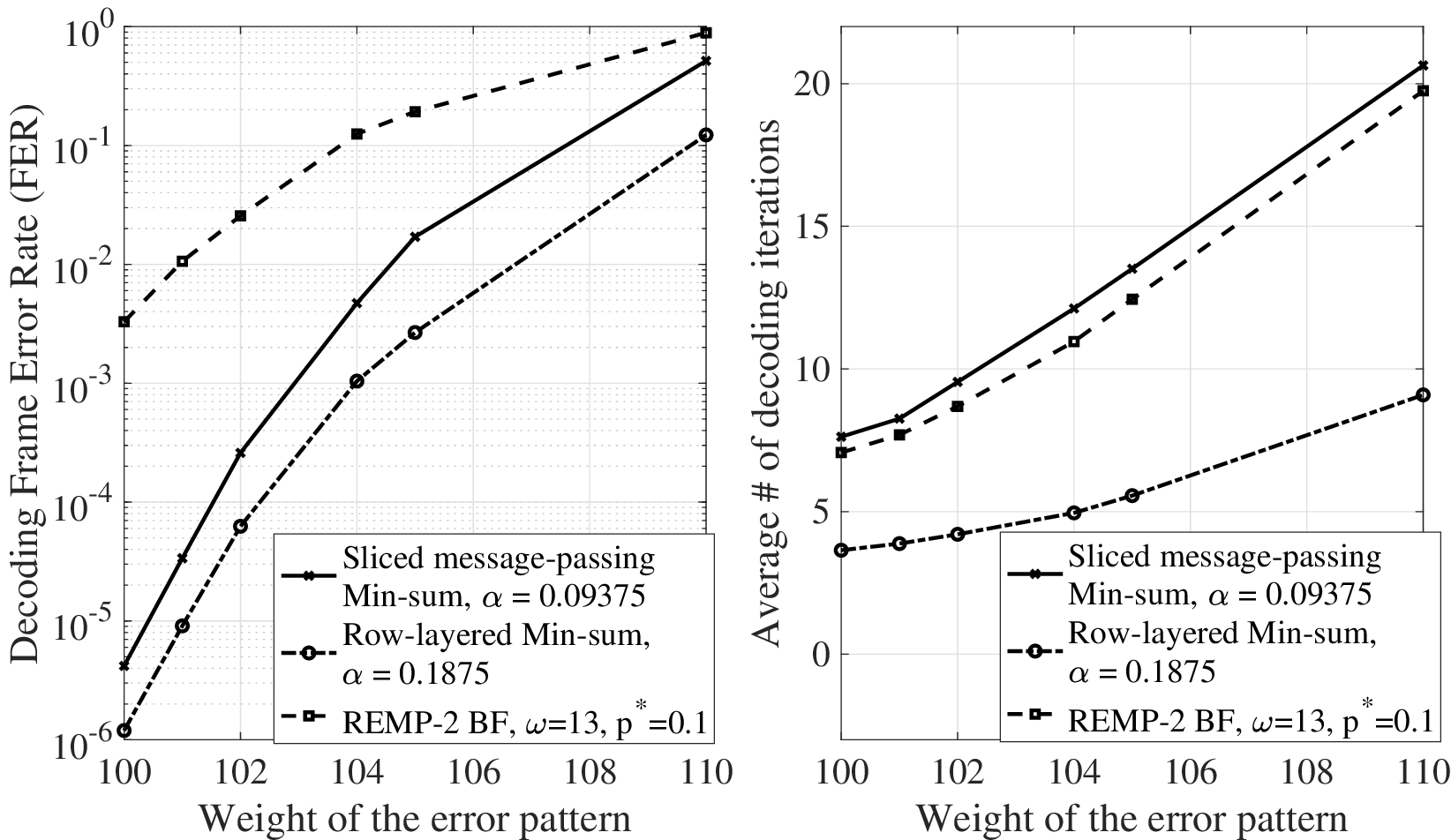}
    \vspace {-0.5em}\caption{(a) Decoding FER and (b) average number of decoding iterations of sliced message-passing Min-sum, row-layered Min-sum, and REMP-2 BF decoding for an MDPC code with $(n_0,r,w)=(2, 4801,45)$ and $I_{\text {max}}=30$.}\label{serial_layered_perf}
    \end{center}
    \vspace {-1.5em}
\end{figure}

Fig. \ref{serial_layered_perf} shows the simulation results of the sliced message-passing \cite{SlicedLDPC} Min-sum decoder that updates the c2v and v2c messages at the end of each iteration and a row-layered decoder that consists of one row in each layer for a randomly generated MDPC code with $(n_0,r,w)=(2, 4801,45)$. The value of the scalar, $\alpha<1$, affects the decoding frame error rate (FER). To simplify the scalar multiplier, the scalar is allowed to have 6 digits in the fractional part with at most two nonzero digits in our design. The scalar leading to the lowest FER is found from simulations for each decoder. Each v2c and c2v message is represented by 4-bit integer magnitude and 1-bit sign. In this case, setting $C$ to $9$ leads to good performance. As shown in Fig. \ref{serial_layered_perf}, row-layered decoding reduces the average number of decoding iterations substantially compared to the sliced message-passing decoder. When the maximum number of decoding iterations is set to a limited number, such as $I_{\text {max}}=30$, the FER of the row-layered decoder is also lower due to the faster convergence. Most previous MDPC decoder designs consider BF algorithms \cite{Liva, BIKE, Baldi,ErrorFloor, GuneysuDATE, Guneysu, Hu} due to their simplicity. For comparison, simulation results of the REMP-2 BF algorithm \cite{Liva} with optimal parameter settings are included in Fig. \ref{serial_layered_perf}. The REMP-2 algorithm has one of the lowest FERs among BF algorithms. However, its performance is still orders of magnitude worse compared to that of the Min-sum algorithm.


\subsection{Parallel MDPC decoders}
\begin{figure}[t] 
    \begin{center}
    \includegraphics[width=.45\textwidth]{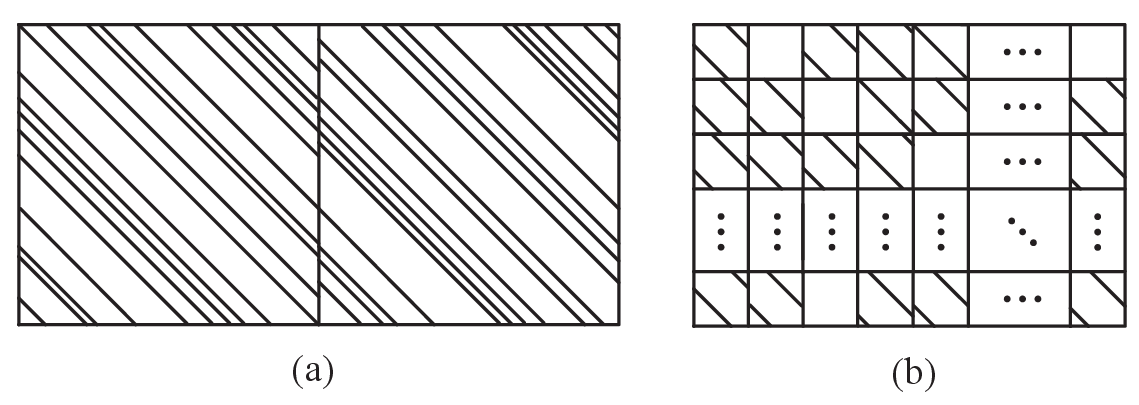}
    \vspace {-0.5em}\caption{Example parity check matrices of (a) an MDPC code used in the McEliece cryptosystem; (b) an LDPC code used for error correction in digital communication and storage systems.}\label{H}
    \end{center}
    \vspace {-1.5em}
\end{figure}

Many hardware implementation architectures have been developed for row-layered decoders of QC-LDPC codes \cite{ParaLayeredLDPCArch1,ParaLayeredLDPCArch2,ParaLayeredLDPCArch3}. The parity check matrices of these codes consist of a large number of smaller zero or CPMs as shown in Fig. \ref{H}(b). In this figure, the nonzero entries are represented by diagonal lines. Since each row or each column has exactly one single nonzero entry in a CPM, efficient parallel QC-LDPC decoders are achieved by processing one or multiple CPMs in each clock cycle. However, the MDPC codes used in the McEliece cryptosystem exhibit  fundamentally different structures in their parity check matrices as illustrated in Fig. \ref{H}(a). Due to the irregular distribution of nonzero entries in the randomly generated $\mathbf H$ matrix, the parallel designs of QC-LDPC decoders are not applicable to MDPC decoders.

In the $L$-parallel Min-sum MDPC decoder of \cite{McElieceSlicedMP}, the $\mathbf H$ matrix is horizontally divided into $L$ segments and the decoder tries to process one entry from each segment simultaneously in each clock cycle. However, due to the unbalanced number of nonzero entries among the $L$ segments, the effective speedup factor of this design is quite limited. From simulations on 1000 randomly generated MDPC codes with $(n_0,r,w)=(2, 4801,45)$, on average, the 2-parallel design in \cite{McElieceSlicedMP} can  achieve 1.82 times speedup compared to the serial design. The effective speedup becomes 3.15 times on average for $4$-parallel decoding, and this design is even less effective for higher parallelism despite the $L$ times more check node units (CNUs) and larger variable node units (VNUs). Furthermore, one memory bank is needed to record the messages for each segment of $\mathbf H$ and its size is decided by the worst-case number of messages to store. Due to the irregularity of the locations of nonzero entries, the overall size of the memories increases for higher parallelism.

\section{Row-Layered Min-Sum MDPC Decoder with Finite Precision}

In this section, a row-layered MDPC Min-sum decoder architecture is first presented to explain the involved computation units and decoding data flow. It serves as a starting point for the design of the parallel decoder to be detailed in Section V. Then a low-complexity method is proposed to mitigate the performance loss caused by finite precision representation of messages.

\subsection{Row-layered MDPC decoder architecture}

\begin{figure}[t] 
    \begin{center}
    \includegraphics[width=.45\textwidth]{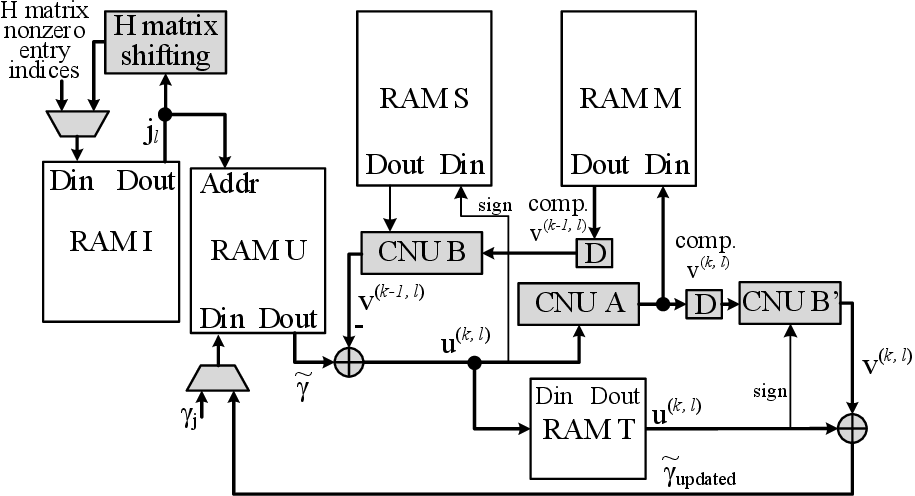}
    \vspace {-0em}\caption{Top-level architecture of row-layered Min-sum MDPC decoder.}\label{serial_overall}
    \end{center}
    \vspace {-0.5em}
\end{figure}

\begin{figure}[t] 
    \begin{center}
    \includegraphics[width=.48\textwidth]{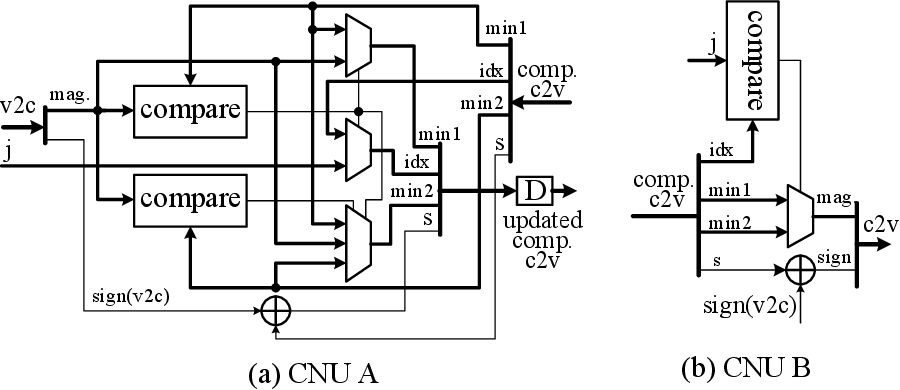}
    \vspace {-0.5em}\caption{(a) Architecture of CNU A; (b) architecture of CNU B.}\label{CNU}
    \end{center}
    \vspace {-1.5em}
\end{figure}

A row-layered Min-sum decoder for MDPC codes can be implemented by extending the architecture for LDPC codes in \cite{MyBook} as shown in Fig. \ref{serial_overall}. In the $\mathbf H$ matrix of an MDPC code for the McEliece cryptosystem, each row is a cyclic shift of the previous row. In order to increase the hardware efficiency and reduce the memory requirement, only the indices of the nonzero entries in one row of the $\mathbf H$ matrix are stored in RAM I. As the indices are read out for the decoding, they are incremented by one mod $r$ to derive the indices of the nonzero entries in the next row of $\mathbf H$ by the ``H matrix shifting'' block. Subsequently, these results are written back to RAM I.

Since the c2v messages are initially zero, the initial value of the {\it a posteriori} information for bit $j$ equals the channel information $\gamma_j$, which is set to $+C$ or $-C$ depending on whether bit $j$ is `0' or `1'. These information are loaded into RAM U in the beginning of the decoding. The {\it a posteriori} information is updated according to \eqref{apost} and written back to RAM U, and the v2c messages are calculated by subtracting the c2v messages of the same check node from the previous iteration from the {\it a posteriori} information as shown in \eqref{v2c}. Hence, the channel information does not need to be stored. 

RAM M is used to hold the c2v messages for all rows in a compressed format containing $min1$, $min2$, $idx$, and $s$. A CNU comprises two parts and their architectures are depicted in Fig. \ref{CNU} \cite{MyBook}. CNU A iteratively computes $min1$, $min2$, $idx$, and $s$ values according to Lines 8-11 of Algorithm \ref{minsumalgo} and CNU B derives the actual c2v message from its compressed form as listed in Lines 13 and 14 of Algorithm \ref{minsumalgo}. The sign bits of the v2c messages are stored in RAM S. RAM T is used to buffer $u^{(k,l)}$ until it is added up with the $\alpha v^{(k,l)}$ calculated by the CNU to update the {\it a posteriori} information.  

\begin{figure}[t] 
    \begin{center}
    \includegraphics[width=.48\textwidth]{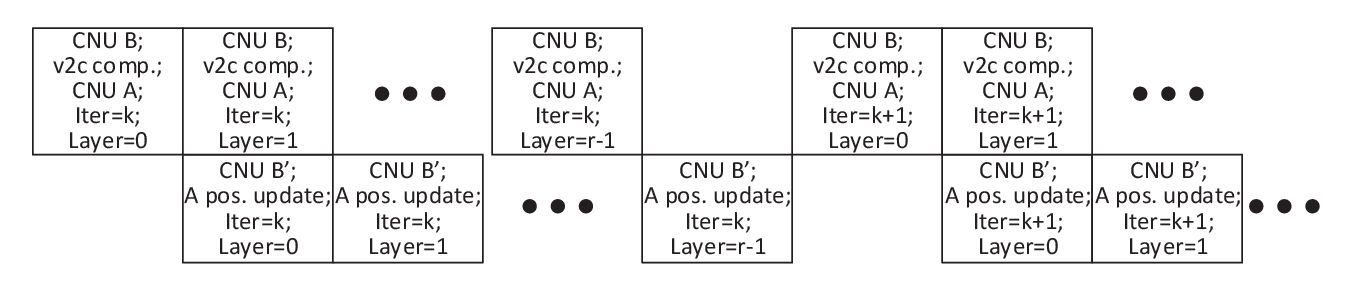}
    \vspace {-0em}\caption{Scheduling of variable and check node processing in the row-layered decoder.}\label{schedule}
    \end{center}
    \vspace {-0.5em}
\end{figure}

During the decoding of a layer, the indices of the nonzero entries of $\mathbf H$ stored in RAM I are read out consecutively and used as the addresses to read out the corresponding $\tilde \gamma$ from RAM U. At the same time, $v^{(k-1,l)}$ are generated by CNU B using the compressed messages stored in RAM M and sign bit stored in RAM S. Then the v2c messages are derived as $u^{(k,l)} =\tilde \gamma - \alpha v^{(k-1,l)}$ and sent to CNU A to compute the $min1$, $min2$, $idx$, and $s$ values as the compressed c2v message for $v^{(k,l)}$. The new sign bits of $u^{(k,l)}$ replace the old sign bits in RAM S. The newly computed compressed $v^{(k,l)}$ overwrite the old values in RAM M and are also sent to CNU B' to derive the actual $v^{(k,l)}$. $\alpha v^{(k,l)}$ added with the $u^{(k,l)}$ read out from RAM T is the updated {\it a posteriori} information, which are written back to RAM U. The addresses for accessing all RAMs except RAM U are generated using counters, and hence are not explicitly shown in Fig. \ref{serial_overall}. Fig. \ref{schedule} shows the scheduling of the computations in the row-layered decoder. The updating of the {\it a posterior} information using the c2v messages generated from the decoding of layer $l$ overlaps with the check node processing of layer $l+1$. 

\subsection{Performance loss caused by finite precision and mitigation}
In hardware implementations, finite precision is necessary to reduce the number of bits used to represent  the messages in order to reduce memory requirement and simplify computation units. Assume that each v2c and c2v message is represented by a $(q+1)$-bit integer with $q$-bit magnitude and 1-bit sign. As shown in Line 16 in Algorithm \ref{minsumalgo}, the {\it a posteriori} information is the sum of the channel information and scaled c2v messages. To prevent overflow, more bits are used to represent the {\it a posteriori} information and the result of subtracting it by $\alpha v^{(k-1,l)}$ according to \eqref{v2c}. The magnitude of the difference is saturated to $2^q-1$ before it is utilized as the v2c message in CNU A. To simplify the scalar multiplications, the scalars considered in our design have at most two nonzero digits, each of which is either $+1$ or $-1$. In the design of \cite{McElieceSlicedMP},  the results of scalar multiplications are brought back to integer representation. Rounding instead of truncation is utilized to reduce the precision loss. 

\begin{figure}[t] 
    \begin{center}
    \includegraphics[width=.48\textwidth]{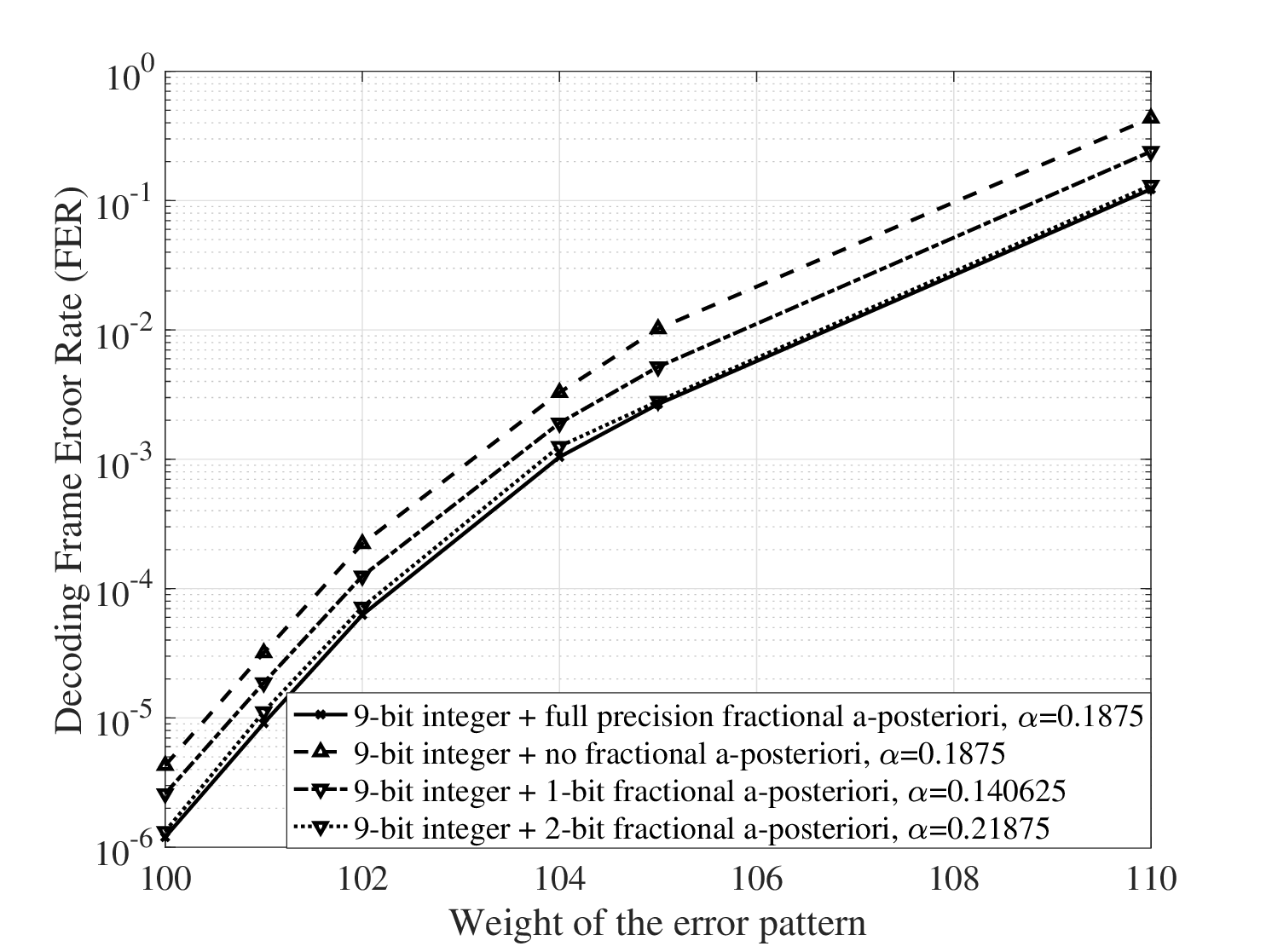}
    \vspace {-0.5em}\caption{FERs of row-layered Min-sum decoding with $4$-bit c2v and v2c message magnitude and different precision of {\it a posteriori} information for an MDPC code with $(n_0,r,w)=(2, 4801,45)$ and $I_{\text {max}}=30$.}\label{mitigation}
    \end{center}
    \vspace {-1.5em}
\end{figure}

Fig. \ref{mitigation} illustrates simulation results for row-layered Min-sum decoding of an MDPC code with $(n_0,r,w)=(2, 4801,45)$, $I_{\max}=30$, and $q=4$ bits representing each c2v and v2c message magnitude. Optimal scalars for each decoding scheme as shown in the figure are derived from simulations and they are small due to the high column weight of the code. Assume that each {\it a posteriori} information is represented by a $(p+1)$-bit integer with $p$-bit magnitude and 1-bit sign. For the scalars shown in Fig. \ref{mitigation} and $w=45$, $p=\lceil log_2(0.21875\times 45\times (2^4-1)+15)\rceil =8$ is sufficient to prevent overflow. However, if the scaled c2v messages are brought back to integer, even with rounding, the row-layered decoding has much higher FER compared to the case where all the fractional bits are kept as shown in Fig. \ref{mitigation}. This is because, following \eqref{apost} and \eqref{v2c}, in row-layered decoding, a single c2v message instead of the sum of c2v messages as in \cite{McElieceSlicedMP} is scaled by a small scalar. The rounding brings an error in the range of (-0.5, 0.5) for each scaled c2v message and the errors are accumulated during the updating of the {\it a posteriori} information.

To mitigate the performance loss, this paper proposes to keep a limited number of fractional bits in the scaled c2v messages and the {\it a posteriori} information to improve their precision. If $i$ fractional bits are kept, the rounding error of each scaled c2v message is reduced to $(-0.5/2^i, 0.5/2^i)$. Fig. \ref{mitigation} shows that keeping two fractional bits in the scaled c2v messages and {\it a posteriori} information can achieve almost the same FER as the decoder with full precision.

\section{MDPC Code Construction Constraints for Parallel Processing}
This section first proposes adding constraints to the construction of the parity check matrices of MDPC codes in order to enable highly efficient parallel processing. Then it is shown that the proposed constraints do not bring noticeable increase in the decoding failure rate or compromise the security of the cryptosystem in a non-negligible way.

\subsection{Constraints on MDPC code construction for parallel processing}
For QC-LDPC codes used in digital communication and storage systems, as shown in Fig. \ref{H} (b), their $\mathbf H$ matrices consist of CPMs or zero matrices of dimension $L\times L$. A row-layered decoder that processes an $L\times L$ block of $\mathbf H$ can be easily designed since there is a single nonzero entry in each row and each column of a CPM. If there is more than one nonzero entries in a row, the CNU will be much more complicated than that in Fig. \ref{CNU}. If there are multiple nonzero entries in a column, then the message updating formula in \eqref{apost} and \eqref{v2c} are not applicable. Formulas for row-layered decoding with two nonzero entries in a column are available in \cite{MultiBlockRow}.

If the $\mathbf H$ matrix of an MDPC code is divided into blocks of size $L\times L$, each block may have 
multiple `1's in a row or column due to the random code construction. Besides, the number of `1's varies among the blocks. The CNUs need to be designed to handle the worst case and the memories storing the messages need to be wide enough to store the maximum number of messages in a block. The v2c message updating formulas become much more complicated for processing more than one nonzero entry at a time. Additionally, many of the hardware units are idling when the decoder is processing a block with a smaller number of nonzero entries. As a result, such parallel MDPC decoders have large area overhead and low efficiency. Constraints need to be added to the $\mathbf H$ matrices to make efficient parallel processing possible. 

To enable efficient parallel decoding, this paper proposes the following constraint on the construction of the $\mathbf H$ matrices of MDPC codes, which are specified by the $r$-bit vectors in the first columns of the $n_0$ sub-matrices. For $L$-parallel processing, it is required that the circular distance between any pair of nonzero entries in each of the $n_0$ $r$-bit vectors is at least $L$. For two entries whose indices are $i$ and $j$, their circular distance is defined as $\min \{|i-j|, r-|i-j|\}$. A matrix satisfying the above constraint is referred to as a "matrix constrained by $L$" in the remainder of this paper. 

\begin{figure}[t] 
    \begin{center}
    \includegraphics[width=.48\textwidth]{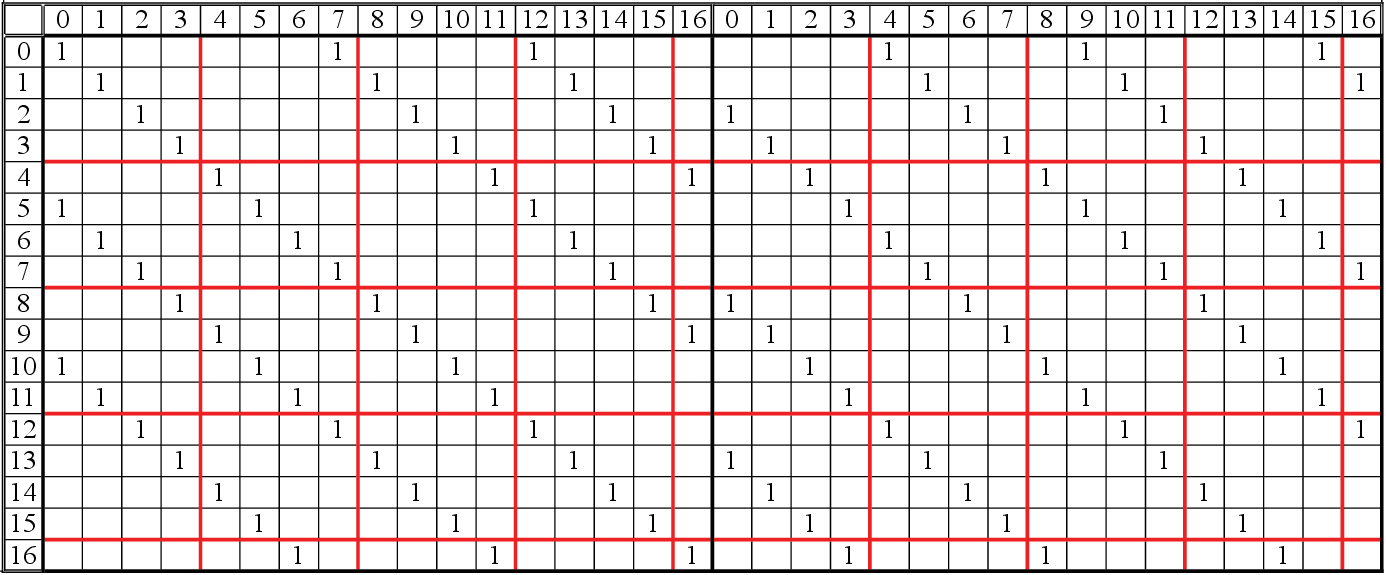}
    \vspace {-0.5em}\caption{Example $\mathbf H$ matrix divided according to fixed indices of a toy MDPC code with $(n_0,r,w)=(2, 17, 3)$ constrained by $L=4$. }\label{division_fixed}
    \end{center}
    \vspace {-1.5em}
\end{figure}

If the proposed constraint is satisfied, then the $\mathbf H$ matrix can be divided into $L \times L$ blocks, where each block has at most a single nonzero entry in each row and each column. One block can be processed in each clock cycle to achieve $L$-parallel processing. Fig. \ref{division_fixed} illustrates an example $\mathbf H$ matrix of a toy MDPC code with $(n_0,r,w)=(2, 17, 3)$ constrained by $L=4$. In this example, the $\mathbf H$ matrix is divided into blocks according to fixed indices. However, it is possible to divide the matrix in alternative ways to achieve even more efficient parallel processing and the details will be presented in Section V. Since the size of the submatrix, $r$, is a prime number, there are some columns and rows by the edges of each submatrix that do not form complete $L \times L$ blocks. These exceptions can be easily handled in the decoder as will be explained in Section V. 

\subsection{Analyses on the effects of the constraint on decoding performance and security}
The $\mathbf H$ matrix was originally randomly constructed for the McEliece cryptosystem. However, our design adds a constraint to the $\mathbf H$ matrix construction in order to enable efficient parallel decoding. The potential effects of the proposed constraint on the number of possible keys, decoding failure rate, and resilience towards reaction attacks are analyzed in this subseciton. 

 \begin{table}[t]
    \renewcommand\arraystretch{1.15}
    \caption{Number of possible private keys for different $L$ derived from simulations on $10^9$ randomly generated submatrices of MDPC codes with $(n_0,r,w)=(2, 4801,45)$.}\label{number_keys} \vspace{-1em}
    \begin{center}
    \begin{tabular}{c|c}
    \hline
    $L$ & number of possible keys \\
    &$((s_1/10^9) \times C(4801, 45))^2$\\\hline\hline
    2 & $((226272570/10^9) \times C(4801, 45))^2 \approx 2^{723}$  \\\hline
    4 & $((98768078/10^9) \times C(4801, 45))^2 \approx 2^{721}$  \\\hline
    8 & $((17880319/10^9) \times C(4801, 45))^2 \approx 2^{716}$  \\\hline
    16 & $((465981/10^9) \times C(4801, 45))^2 \approx 2^{705}$\\\hline
    32 & $((119/10^9) \times C(4801, 45))^2 \approx 2^{681}$ \\\hline
    \end{tabular}
    \end{center}
\end{table}

As mentioned previously, the private key of the MDPC-based scheme consists of $n_0$ $r$-bit vectors specifying $\mathbf H_0, \mathbf H_1,\cdots, \mathbf H_{n_0-1}$, and each $r$-bit vector has weight $w$. $\mathbf  H_{n_0-1}$ needs to be invertible. However, the possibility of a randomly generated $r$-bit vector with weight $w$ leading to non-invertible $\mathbf  H_{n_0-1}$ is negligible \cite{MDPCMcEliece}. Hence the total number of possible private keys can be estimated as $(C(r, w))^{n_0}$. It is very difficult to derive a closed-form formula for the number of possible $r$-bit vectors when the proposed constraint is added. Hence, simulations were carried out to estimate the total number of possible private keys satisfying the proposed constraint. A large number of $r$-bit vectors with weight $w$ are randomly generated. Then the vectors satisfying the proposed constraints are counted. Denote these two numbers by $s_0$ and $s_1$, respectively. Then the total number of possible private keys satisfying the proposed constraint can be estimated as $((s_1/s_0)\times C(r, w))^{n_0}$. For MDPC codes with $(n_0,r,w)=(2, 4801,45)$, the total number of possible $4801$-bit vectors with weight 45 is $C(r,w) = C(4801,45)=3.1 \times 10^{109}$. Hence, the total number of possible keys is $(3.1 \times 10^{109})^2 \approx 2^{727}$ originally. From $10^{9}$ $4801$-bit vectors with weight 45 randomly generated from simulations, $119$ of them satisfy the proposed constraint with $L=32$. Therefore, the total number of keys under our proposed constraint with $L=32$ is $((119/10^9)\times 3.1 \times 10^{109})^2 \approx 2^{681}$. Although the number of possible keys is reduced, it is still very large so that the private key is not recoverable by exhaustive search. For different $L$, the numbers of possible private keys are listed in Table \ref{number_keys}. As shown in Table \ref{para}, the MDPC codes considered for higher security level have lower density. As a result, the number of possible keys is reduced by a smaller percentage when the proposed constraint is reinforced.

\begin{figure}[t] 
    \begin{center}
    \includegraphics[width=.48\textwidth]{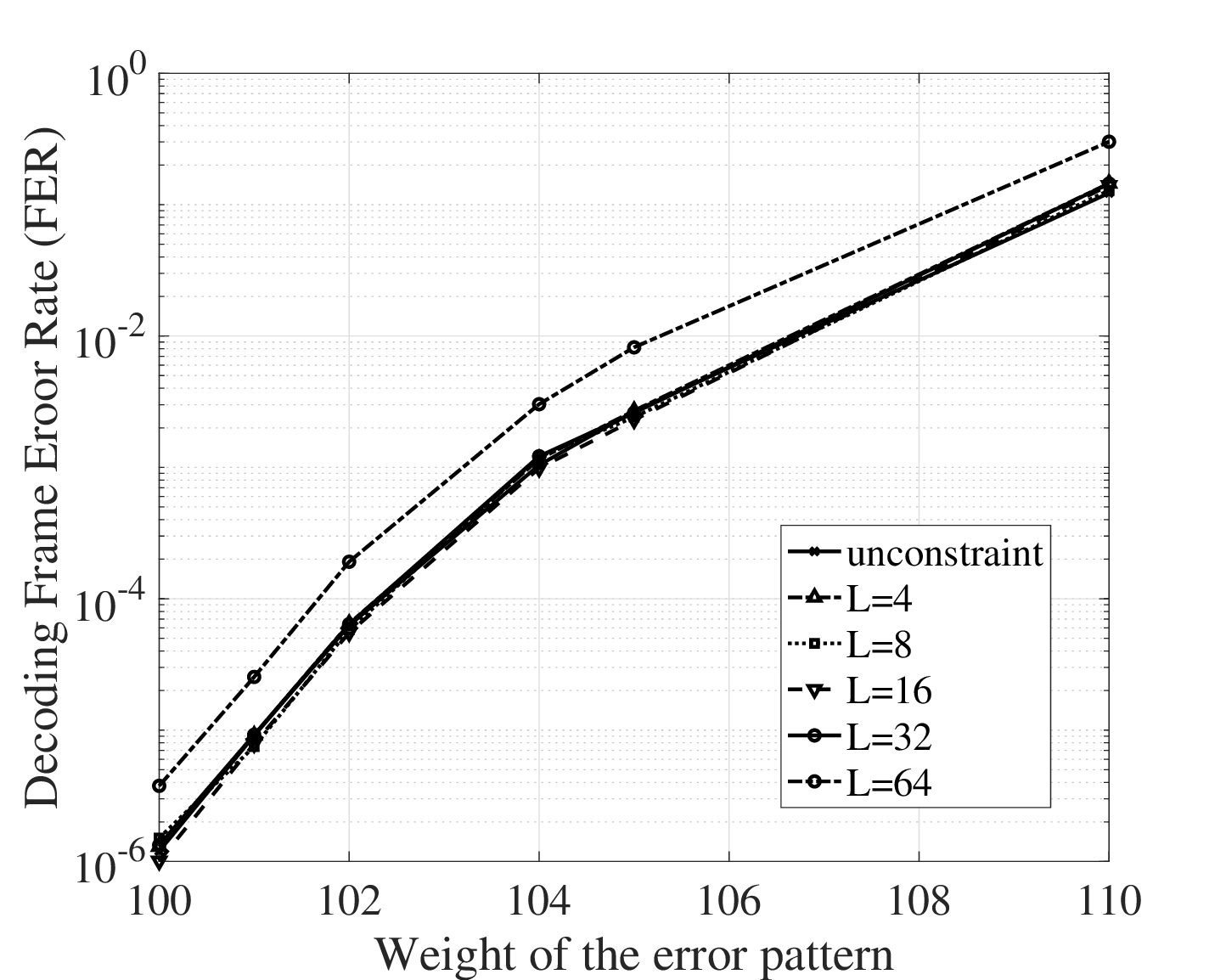}
    \vspace {-0.5em}\caption{FERs of Min-sum row-layered decoding for MDPC codes with $(n_0,r,w)=(2, 4801,45)$ and $I_{\text {max}} = 30$ without constraint and constrained by different $L$.}\label{constraint_FER}
    \end{center}
    \vspace {-1.5em}
\end{figure}

The proposed constraint may affect the decoding failure rates of MDPC codes. To investigate the possible effects, simulations have been carried out to find the FERs of random MDPC codes with $(n_0,r,w)=(2, 4801,45)$ constrained by different $L \le 64$. For each $L$, more than $10$ random codes have been generated. It is found from simulations that different codes with the same $L\leq 32$ have very similar FERs. Hence only one code is chosen for each $L$ to show the FER in Fig. \ref{constraint_FER}. From this figure, it is observed that the FER curves for different $L$ $\le 32$ almost overlap with that of the code without the proposed constraint, indicating that the proposed constraint does not lead to any performance loss when the parallelism is moderate or small. However, when $L=64$, FER increase is observed for one of the randomly generated codes and this is shown in Fig. \ref{constraint_FER}. One possible reason is that, for large $L$, the number of possible locations of the nonzero entries in the $\mathbf H$ matrix is significantly reduced, which may lead to more $4$-cycles. The number of possible nonzero entry locations increases with $r$, and hence longer MDPC codes may tolerate larger $L$.

Various attacks targeting at the McEliece cryptosystem are available in the literature. The reaction attacks \cite{Reaction,BaldiAttack} that try to recover the secret $\mathbf{H}$ matrix by utilizing decoding failures are among the most powerful. The effects of the proposed $\mathbf{H}$ matrix construction constraint on the resiliency to the reaction attacks need to be evaluated. In our analysis, the key-recovery GJS attack \cite{Reaction} is considered since it is one of the most critical reaction attacks. This attack consists of two steps: distance spectrum construction and private key reconstruction. Denote the first row of $\mathbf H_0$ by $\mathbf h_0$. The distance spectrum of $\mathbf h_0$, $D(\mathbf h_0)$, is a list of the circular distance $d\le \lfloor r/2 \rfloor$ between any pair of nonzero entries in the vector. To construct $D(\mathbf h_0)$, for every $1\leq d\le \lfloor r/2 \rfloor$, $M$ $n_0r$-bit vectors, each with $t$ `1's  arranged as $\lfloor t/2 \rfloor$ pairs with circular distance $d$ within the first $r$ bits of the vector are sent to the decoder, and the corresponding FER is recorded. $M$ is a large integer. The FER is lower for those $d \in D(\mathbf h_0)$, and accordingly the distance spectrum is constructed. The second step reconstructs the $\mathbf H$ matrix from $D(\mathbf h_0)$. From calculations and simulations in \cite{Reaction}, the complexity of the GJS attack is dominated by the first step. Hence, the analysis on the impacts of the proposed constraint only needs to be carried out on the first step.

The number of decoding trials in the first step of the GJS attack for codes without the proposed constraint is $\lfloor r/2 \rfloor \cdot M$. $M$ needs to be large enough to collect a non-trivial number of decoding failures and hence it is decided by the decoding failure rates of MDPC codes. When the parallelism is moderate or small, the proposed constraint does not lead to any FER degradation. Hence, the same $M$ is needed to attack the codes adopting our constraint. On the other hand, since $1\leq d<L$ are not possible entries in the spectrum, the total number of decoding trials to attack the code with our proposed constraint is reduced to $(\lfloor r/2 \rfloor-L) \cdot M$. $r$ is a very large number as shown in Table \ref{para}. Therefore, the number of decoding trails is only reduced by a small proportion for our scheme. For an example MDPC code with $(n_0,r,w)=(2, 4801,45)$ constrained by parallelism $L=32$, the number of decoding trials is only reduced by $32/2400 = 1.33\%$. For MDPC codes considered for higher security levels, due to the larger $r$, the number of decoding trails is reduced by an even smaller percentage. Thus, it can be concluded that the proposed constraint on MDPC code construction only negligibly increases the vulnerability towards the GJS attack.

\section{Highly Efficient Parallel MDPC Decoder Architecture}

In this section, a dynamic scheme is proposed to divide the $\mathbf H$ matrix into $L\times L$ identity blocks to achieve $L$ times speedup in $L$-parallel processing. Then an even-odd message storage method is developed to enable the proposed matrix division in order to realize a highly efficient parallel MDPC decoder.

\subsection{Dynamic $\mathbf H$ matrix division scheme}

In Fig. \ref{division_fixed}, the $\mathbf H$ matrix is divided into $L\times L$ blocks in a straightforward manner according to fixed indices. In this case, although the decoder processes an $L\times L$ block in each clock cycle, the achievable speedup is far less than $L$ since a large portion of the blocks have less than $L$ nonzero entries due to the irregular distribution of the nonzero entries in the $\mathbf H$ matrices MDPC codes. From simulations on $10^9$ randomly generated submatrices of MDPC codes with $(n_0,r,w)=(2, 4801,45)$, Table \ref{speedup_simu} shows the average number of nonzero entries in those $L\times L$ nonzero blocks for different $L$. The achievable speedup compared to a serial design is only around 50-60\% of $L$ for larger $L$.

\begin{table}[t]
    \renewcommand\arraystretch{1.15}
    \caption{Average numbers of nonzero entries in $L\times L$ nonzero blocks of $\mathbf H$ matrices from $10^9$ randomly generated submatrices of MDPC codes with $(n_0,r,w)=(2, 4801,45)$. }\label{speedup_simu} \vspace{-1em} 
    \begin{center}
    \begin{tabular}{c|c|c}
    \hline
    $L$ & \multicolumn{2} {c}{ Avg. \# of nonzero entries in $L\times L$ blocks} \\\cline{2-3}
    &straightforward matrix division & dynamic matrix division \\
&  & (proposed)\\\hline\hline
    2 & 1.346 &2 \\\hline
    4 & 2.319 &4\\\hline
    8 & 4.373 &8\\\hline
    16 & 8.643 &16\\\hline
    32 & 18.103 &32\\\hline
    \end{tabular}
    \end{center}
\end{table}

\begin{figure}[t] 
    \begin{center}
    \includegraphics[width=.48\textwidth]{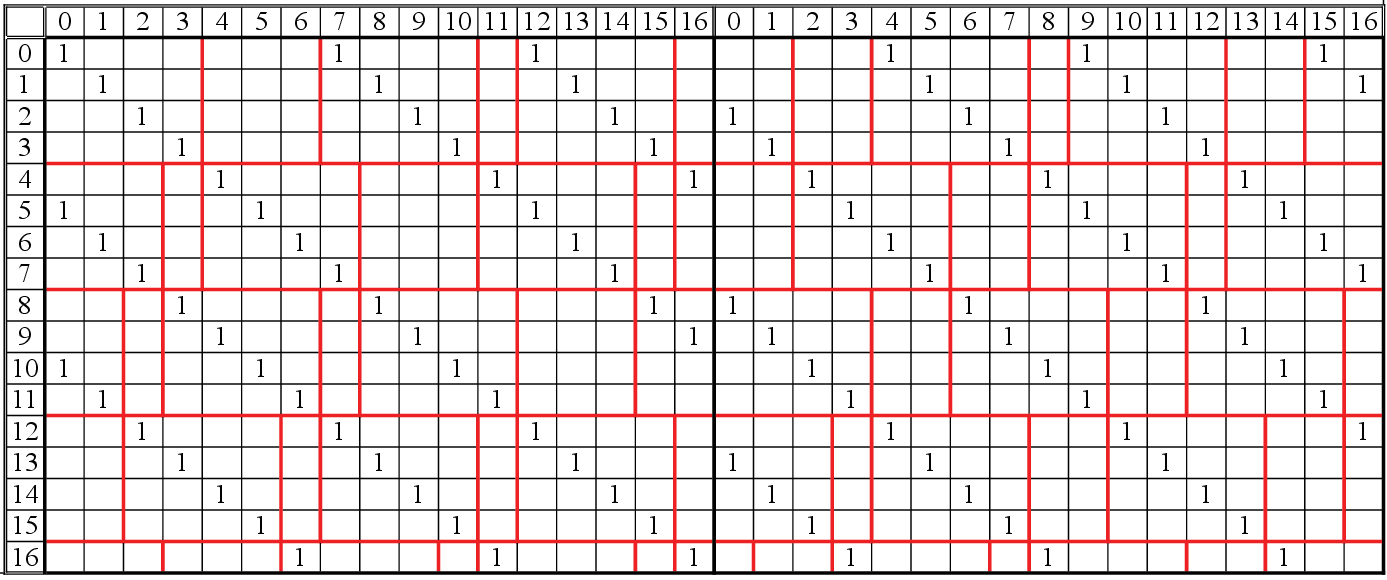}
    \vspace {-0.5em}\caption{Example $\mathbf H$ matrix divided by the proposed dynamic scheme for a toy MDPC code with $(n_0,r,w)=(2, 17, 3)$ constrained by $L=4$. }\label{division_identidy}
    \end{center}
    \vspace {-0.5em}
\end{figure}

To achieve $L$ times effective speedup by processing an $L\times L$ block each time, this section proposes to divide the $\mathbf H$ matrix dynamically into $L\times L$ identity blocks. An example $\mathbf H$ matrix divided by the proposed scheme for $L=4$ is illustrated in Fig. \ref{division_identidy}. The top left corner of each $L\times L$ identity block is a nonzero entry in the rows of $\mathbf H$ whose indices are integer multiples of $L$. Due to the cyclical shift property of $\mathbf H$, some identity blocks wrap around the column edges of the submatrices, such as the identity formed by columns 15,16,0,1 in the first 4 rows of the second submatrix shown in Fig. \ref{division_identidy}. Although $r$ is a prime number and hence is not divisible by $L$, the last $r-\lfloor r/L\rfloor L$ rows of $\mathbf H$ are divided into blocks in the same way, except that they are incomplete identity blocks. They can still be processed by the same hardware units for the full identity blocks by disabling the units for the incomplete rows.

Using the proposed dynamic $\mathbf H$ matrix division scheme, the row-layered decoder can process one block of $L$ rows as a layer. In each layer, one identity block is processed in each clock cycle. Since each of the identity blocks has exactly $L$ nonzero entries, except those in the last $r-\lfloor r/L\rfloor L$ rows of $\mathbf H$, the decoder using the proposed matrix division can achieve almost $L$ times speedup compared to the serial design as shown in Table \ref{speedup_simu}.

\subsection{Highly efficient parallel row-layered decoder architecture}

\begin{figure}[t] 
    \begin{center}
    \includegraphics[width=.48\textwidth]{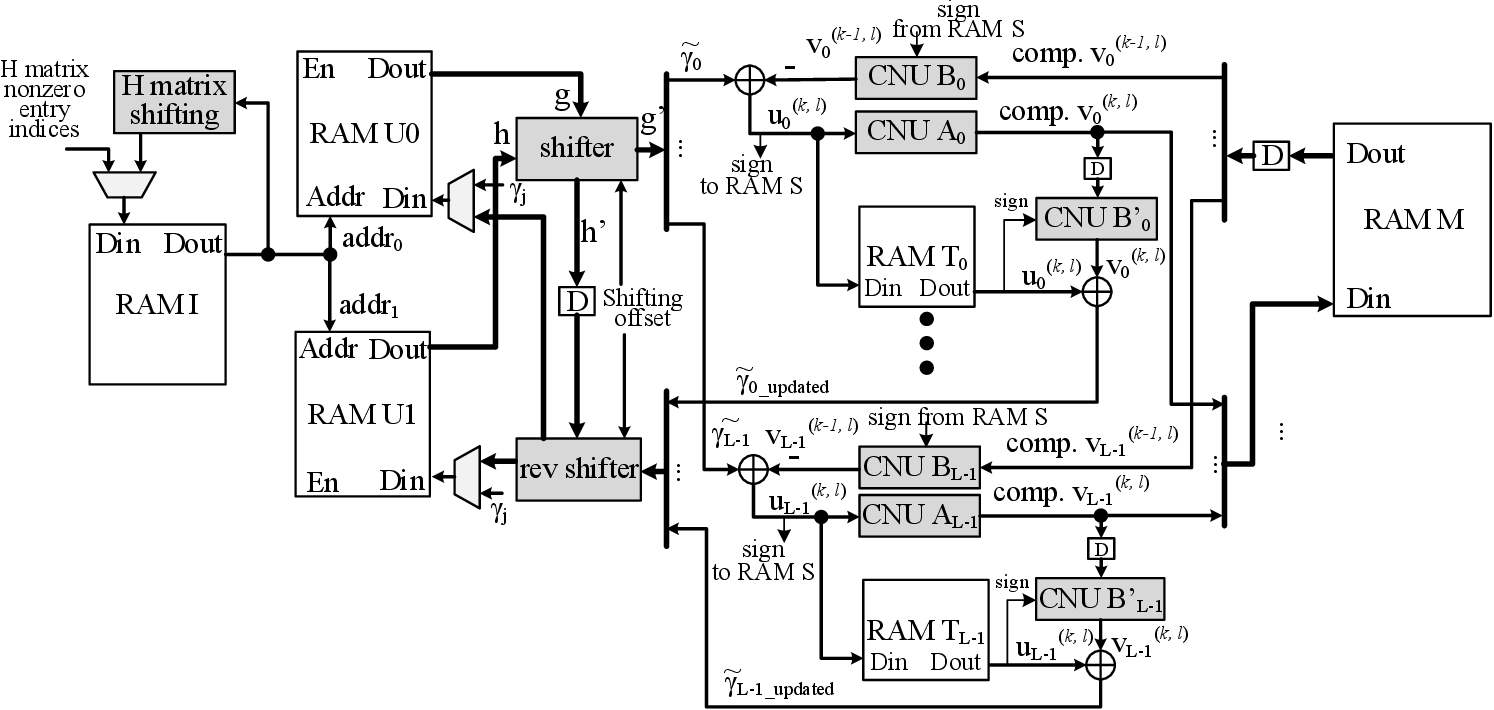}
    \vspace {-0.5em}\caption{Top-level architecture of the proposed $L$-parallel row-layered Min-sum MDPC decoder. }\label{parallel_overallarch}
    \end{center}
    \vspace {-1.5em}
\end{figure}

The top-level block diagram of the proposed $L$-parallel row-layered Min-sum MDPC decoder is shown in Fig. \ref{parallel_overallarch}. In our design, only the nonzero identity blocks of $\mathbf H$ are processed in order to reduce the latency and increase the hardware utilization efficiency. To reduce the memory requirement, only the column indices for the `1's in the top left corners of the identity blocks in the first layer are stored into RAM I in the beginning of the decoding. For example, the column indices recorded for the first layer of the first submatrix in the example matrix in Fig. \ref{division_identidy} are 0,7,12. During decoding, the column indices for the next layer are derived from those of the current layer by adding  $L$ mod $r$ in the ``H matrix shifting" block and the results are written back to RAM I. 

RAM U is the memory used for storing the most updated {\it a posteriori} information for each decoder input bit and it is initialized with the channel information. For an $L$-parallel decoder, the {\it a posteriori} information of $L$ consecutive bits are stored in one address of RAM U. However, an identity block of $\mathbf H$ can start from any column index in the proposed dynamic matrix division scheme. Hence, the information belonging to an identity block may not be in one address of RAM U. However, they can always be found from two consecutive addresses of RAM U. To enable the simultaneous access of two consecutive addresses, RAM U is divided into two banks: RAM U0 storing the {\it a posteriori} information for the even block columns of $\mathbf H$ and RAM U1 recording those of the odd block columns.

\begin{figure}[t] 
    \begin{center}
    \includegraphics[width=.48\textwidth]{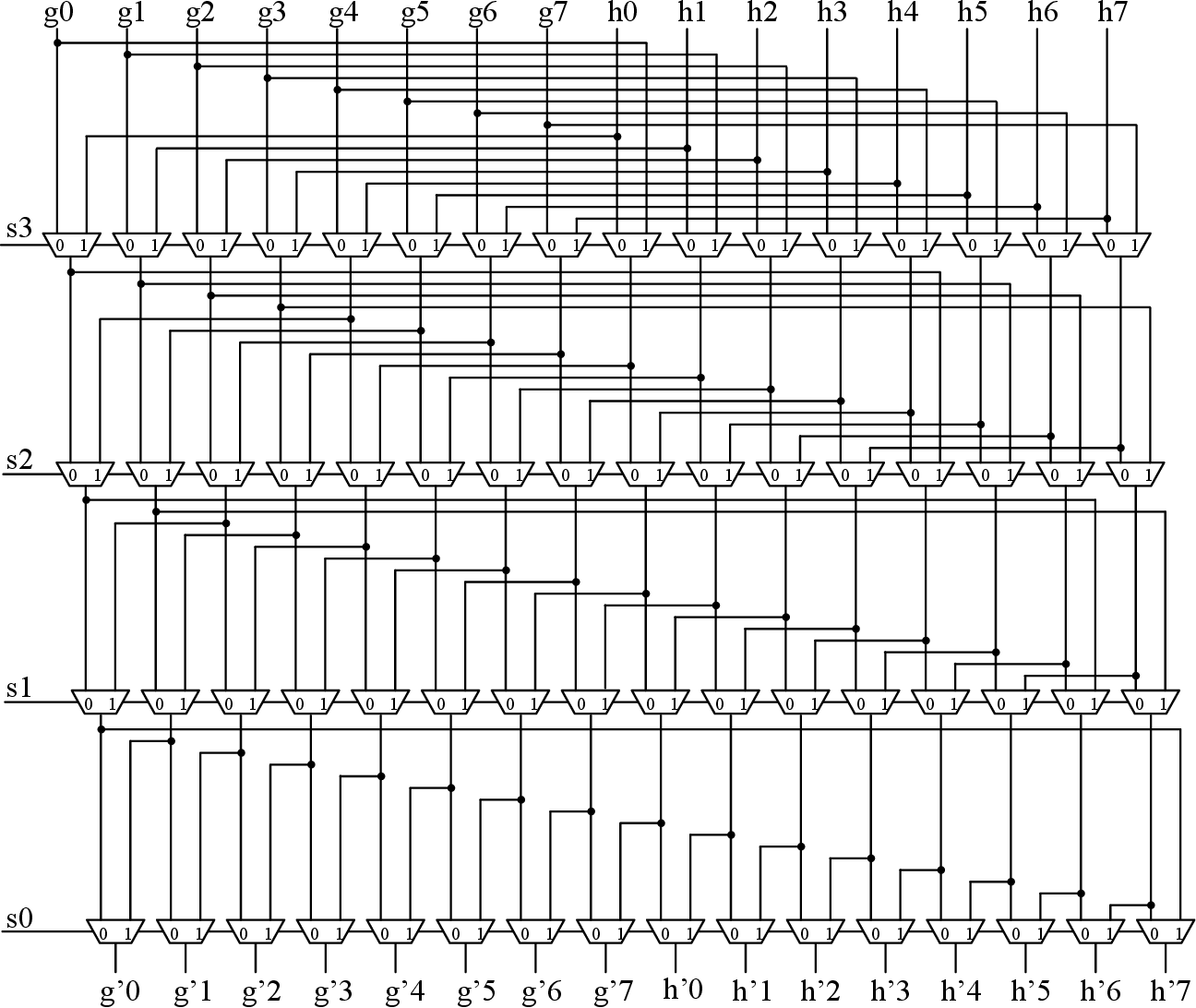}
    \vspace {-0em}\caption{The architecture of the shifter for an example $L=8$.}\label{shifter}
    \end{center}
    \vspace {-0.5em}
\end{figure}

Take an $L=32$-parallel decoder for an MDPC code with $(n_0,r,w)=(2, 4801,45)$ as an example. A column index is represented by $\lceil log_2(2\times 4801)\rceil =14$ bits as $a= a_{13}a_{12}\cdots a_{0}$. The address for accessing RAM U1 is the higher $14-\lceil log_2 2L\rceil$=8 bits, $a_{13}a_{12}\cdots a_{6}$, and that for accessing RAM U0 is $a_{13}a_{12}\cdots a_{6} +a_{5}$. The lower $\lceil\log_2 2L\rceil=6$ bits of $a$ decide which of the $L$ entries to take from the $2L$ information read from RAM U0 and U1. This is implemented by the shifter shown in Fig. \ref{shifter}. To simplify the illustration, this architecture is for an example case of $L=8$. The outputs of RAM U0 and U1 are denoted by $g_0, g_1,\cdots, g_7$ and $h_0,h_1, \cdots, h_7$, respectively, in Fig. \ref{shifter}. Each of them has $p+1$ bits. This shifter consists of $\lceil \log_2 (2L)\rceil$ levels of multiplexors. The signal $s_3s_2s_1s_0=a_3a_2a_1a_0$ decides the number of locations cyclically shifted to the left. At the output of the shifter, $g'_0,g'_1,\cdots, g'_7$ are the $L$ consecutive information starting from index $a$. After these {\it a posteriori} information are updated in the decoder, they are sent to the reverse shifter with $h'_0,h'_1,\cdots, h'_7$, and the outputs are written back to the same addresses at RAM U0 and U1.



Similar to that in the serial decoder introduced in Section III, the RAM M in our proposed parallel decoder is used to hold the c2v messages in a compressed format containing $min1$, $min2$, $idx$, and $s$ for each $\mathbf H$ matrix row. However, since $L$ rows of $\mathbf H$ are processed at a time, the compressed messages for $L$ rows are stored in each address of RAM M in Fig. \ref{parallel_overallarch}. A RAM T serves as a buffer for all the v2c messages during the processing of one row. Since an $L$-parallel design processes $L$ rows simultaneously, $L$ copies of RAM T are needed. 

\newcolumntype{K}{>{\centering\arraybackslash}m{0.6cm}}
\newcolumntype{L}{>{\centering\arraybackslash}m{2.3cm}}
\newcolumntype{M}{>{\centering\arraybackslash}m{2.8cm}}
\begin{table}[!t]
    \renewcommand\arraystretch{1.25}
    \caption{Sizes of the RAM banks utilized in the proposed $L$-parallel decoder for $(n_0,r,w)$ MDPC codes with $q$-bit c2v and v2c message magnitude, and $p$-bit {\it a posteriori} information magnitude.}\label{RAMsize} \vspace{-1em}
    \begin{center} 
    \begin{tabular}{@{}c|K|L|M@{}}
    \hline
    RAM name & \# of banks & size of each bank & values stored\\\hline\hline
    RAM I & $1$ & $ (n_0w) \times \lceil \log_2 (n_0r)\rceil $ &column indices of nonzero entries in one row of the $\mathbf H$ matrix\\\hline
    RAM M & $1$ & $\lceil r/L\rceil \times (L (2q+1+\lceil log_2 (n_0r)\rceil))$ & the most updated compressed version of all $v_{i,j}$\\\hline
    RAM S & $1$ & $ (n_0w \lceil (r/L)\rceil) \times L$ & sign bits for all $u_{i,j}$ in previous decoding iteration\\\hline
    RAM U & $2$ & $(n_0\lfloor r/(2L) \rfloor) \times (L(p+1)) $ & the most updated {\it a posteriori} information for all variable nodes\\\hline
    RAM T &$L$ & $(n_0w) \times (p+1)$& $u_{i,j}$ for one row of $\mathbf H$ during variable node processing\\\hline 
    \end{tabular}
    \end{center}
\end{table}

Table \ref{RAMsize} summarizes the sizes of the RAM banks utilized in the proposed $L$-parallel row-layered decoder for MDPC codes with code parameters $(n_0,r, w)$, $q$-bit c2v and v2c message magnitude, and $p$-bit {\it a posteriori} information magnitude. To support codes with varying parameters, the size of each RAM bank should be configured to its maximum possible value. The contents of each memory, excluding RAM U banks, are accessed consecutively by using a counter to generate the addresses. The addresses used to access RAM U banks are retrieved from RAM I.


In our proposed $L$-parallel decoder, $L$ copies of CNUs are employed. As it was mentioned before, the last block row of $\mathbf H$ has less than $L$ rows. This can be taken care of by disabling the extra CNUs when processing the last block row. For the identity blocks that wrap around the edges of the submatrices of $\mathbf H$, such as the last identity block in the second block row of the first submatrix in Fig. \ref{division_identidy}, the corresponding {\it a posterior} information might be stored in two addresses of the same RAM U bank. To avoid memory access conflicts, registers are used to store the {\it a posterior} information of the last block column for each submatrix in our design. 

\section{Hardware Complexity Analyses} 

\begin{table}[t]
    \renewcommand\arraystretch{1.15}
   \caption{Complexity comparisons for the proposed row-layered Min-sum decoder, the sliced message-passing Min-sum decoder \cite{McElieceSlicedMP}, and the REMP-2 BF decoder \cite{Liva} with $L=2$ for an MDPC code with $(n_0,r,w)=(2, 4801,45)$.}\label{Overall} \vspace{-1em}
   \begin{center}
    \begin{tabular}{c||c|c|c}
    \hline
 & REMP-2 & Sliced message-  & Proposed row- \\
 & BF \cite{Liva}& passing \cite{McElieceSlicedMP}& layered \\
\hline\hline
 RAM I (bits) & 1716 & 1716& 1260 \\\hline
 RAM C (bits) & 19204 & 19204& 0 \\\hline
 RAM M (bits) & 163268 & 220892& 110446 \\\hline
 RAM S (bits) & 633732 & 633732& 432180 \\\hline
 RAM U (bits) & 0 & 0& 105600 \\\hline
 RAM T (bits) & 132 & 330& 1980  \\\hline
 Total memory (bits) &818052 &875874 & 651466 \\
 (normalized) & (0.93) & (1) & (0.74) \\\hline
Logic Area (\# of NAND2) & 1495 &2885 & 4470 \\\hline
 Avg. \# of dec. iter. & 4.41 & 4.66  & 2.05\\\hline
 \# of clks /iter. (worst case)&316866 & 316866&216090  \\\hline
 Max. clock freq. (GHz) & 6.024& 4.132& 4.132 \\\hline
 Avg. latency (ms) & 0.232& 0.357& 0.107\\
 (normalized) & (0.65)& (1) & (0.30)\\ \hline
    \end{tabular}
    \end{center}
\end{table}

This section provides detailed hardware complexity analyses of the proposed decoder for an example $(n_0,r,w)=(2, 4801,45)$ MDPC code. Comparisons with the sliced message-passing decoder \cite{McElieceSlicedMP} and the REMP-2 BF decoder \cite{Liva} are also carried out. For the design in \cite{McElieceSlicedMP}, the memory overhead caused by parallel processing increases rapidly and the speed improvement becomes less significant for larger $L$. Hence $L=2$ designs are compared. 

The logic parts of the design, including VNUs, CNUs, shifters, registers, etc., are synthesized using the GlobalFoundries 22FDX process. The equivalent number of NAND2 gates under a tight 242ps timing constraint is listed in Table \ref{Overall}. On the other hand, the c2v messages in the REMP-2 BF decoder have single-bit magnitudes. Hence the accumulation of the c2v messages in the VNU has shorter data path and it can achieve higher clock frequency. For the $(n_0,r,w)=(2, 4801,45)$ MDPC code, the targeted $t$ is 84 as listed in Table \ref{para}. The average number of decoding iterations listed in Table \ref{Overall} is collected from simulations over 10,000 samples under this parameter. The sizes of the memories for the proposed row-layered Min-sum decoder are calculated from Table \ref{RAMsize} and those for the REMP-2 and sliced message-passing Min-sum decoders have been analyzed in \cite{McElieceSlicedMP}.


Although the proposed decoder requires larger area for the shifter and registers compared to the sliced message-passing decoder in \cite{McElieceSlicedMP}, the memories contribute to the majority of the overall area. Table \ref{Overall} shows that the proposed row-layered decoder achieves $26\%$ total memory requirement reduction compared to that of the sliced message-passing decoder \cite{McElieceSlicedMP}. RAM C is used to store the channel information in the sliced message-passing decoder. However, in the proposed row-layered decoder, the channel information is incorporated in the updated {\it a posteriori} information according to \eqref{apost} after the initialization. Hence, it does not need to be stored in the proposed design and RAM C is eliminated. In the sliced message-passing decoder, two copies of RAM M are needed for storing the c2v messages for the previous decoding iteration and calculating the c2v messages for the current iteration. On the other hand, in the proposed row-layered design, the decoder processes one block row after another. Hence the min1, min2, {\it etc.} for one block row can be stored in registers and one copy of RAM M is needed to record the c2v messages. Lastly, the proposed design requires much smaller RAM S and RAM I because they are not affected by the uneven distribution of nonzero entries in $\mathbf H$. The sliced message-passing decoder in \cite{McElieceSlicedMP} divides the $\mathbf H$ matrix into segments of rows. The number of entries in each RAM S and RAM I bank is more than $w/L$, especially when $L$ is larger, due to the irregularity of $\mathbf H$. This is also the reason that the sliced message-passing decoder requires many more clock cycles per decoding iteration. Besides, the proposed row-layered decoder reduces the number of decoding iterations significantly. As a result, it achieves 70\% latency reduction. Table \ref{Overall} also shows that the proposed decoder has much lower complexity and latency than the REMP-2 decoder in addition to the advantage of several orders of magnitude lower FER as shown in Fig. \ref{serial_layered_perf} (a).

\begin{table}[!t]
    \renewcommand\arraystretch{1.15}
   \caption{Memory requirement comparisons for the proposed parallel row-layered decoder with various $L$ for an MDPC code with $(n_0,r,w)=(2, 4801,45)$.}\label{higherpara} \vspace{-1em}
   \begin{center}
    \begin{tabular}{c||c|c|c|c}
    \hline
 & $L=2$ & $L=8$& $L=16$& $L=32$ \\\hline\hline
 RAM I (bits) & 1260& 1260&1260&1260 \\\hline
 RAM M (bits) & 110446&110584&110768& 111136 \\\hline
 RAM S (bits) & 432180&432720&433440& 434880 \\\hline
 RAM U (bits) & 105600&105600&105600& 105600 \\\hline
 RAM T (bits) & 1980&7920&15840& 31680  \\\hline
 Total memory (bits) &651466 &658084&666908& 684556 \\
 (normalized) & (1) &(1.01)&(1.02)& (1.05) \\\hline
 speedup factor & 2 &8&16&32 \\\hline
    \end{tabular}
    \end{center}
\end{table}

Even for decoders with higher parallelisms, the overall area requirement is dominated by the memories. To show how the complexity of the proposed design grows with the parallelism, Table \ref{higherpara} lists the memory requirements of the proposed decoder for $L=2$, $8$, $16$, and $32$ calculated from Table \ref{RAMsize}. The size of RAM I does not change with the parallelism because it stores the location information of the identity blocks in one layer of $\mathbf H$, which remains the same for different $L$. Since $r$ is a prime number and is not divisible by $L$, the sizes of RAM M, RAM S, and RAM U may change slightly with $L$ due to the ceiling or floor function of $r/L$ as shown in Table \ref{RAMsize}. The total size of RAM T increases linearly with $L$. However, they are much smaller than other memories. As a result, the total size of the memories of the proposed decoder increases very slightly with $L$. Even for a $32$-parallel design, the memories are only $5.1\%$ larger than those of the $2$-parallel decoder. Due to the proposed dynamic $\mathbf H$ matrix division scheme, our proposed $L$-parallel decoder can achieve $L$ times speedup compared to a serial design, ignoring that the last block row of $\mathbf H$ has less than $L$ rows. On the other hand, the speedup achievable by the previous decoder in \cite {McElieceSlicedMP} is far less than $L$, especially for larger $L$, as shown in Table \ref{speedup_simu}.

\section{Conclusions}
For the first time, this paper investigates row-layered Min-sum decoding for MDPC codes with high column weight used in the McEliece cryptosystem. The performance loss caused by scaling and rounding single c2v messages with finite precision in the case of high column weight is identified. Increasing the precision of the {\it a posteriori} information is proposed to compensate the performance loss. Constraints are proposed for the code construction to enable highly efficient parallel decoding. Analyses have proved that the proposed constraints only bring negligible increase on the decoding failure rate and do not compromise the security of the cryptosystem in a non-negligible way. Additionally, a dynamic parity check matrix division scheme and the corresponding parallel decoder architecture are designed to achieve $L$ times speedup for $L$-parallel processing. Future work will focus on further reducing the memory requirement of the decoder.


\begin{thebibliography}{1}
\bibliographystyle{IEEEtran}

\bibitem{McElieceNIST}
D. J. Bernstein, {\it et al.}, (Nov. 2017). ``Classic McEliece: conservative code-based cryptography''. NIST. Accessed: Sep. 2022. [Online]. Available: https://classic.mceliece.org/nist.html.

\bibitem{MDPCMcEliece}
R. Misoczki, J.-P. Tillich, N. Sendrier, and P. S. L. M. Barreto, ``MDPC- McEliece: new McEliece variants from moderate density parity-check codes,'' {\it Proc. IEEE Intl. Symp. on Info. Theory}, pp. 2069-2073, Oct. 2013.

\bibitem{ParaLayeredLDPCArch1}
J. Kim, J. Cho, and W. Sung, ``A high-speed layered min-sum LDPC decoder for error correction of NAND Flash memories,'' {\it Proc. IEEE Intl. Midwest Symp. Circuits and Syst.}, pp. 1-4, Aug 2011.

\bibitem{ParaLayeredLDPCArch2}
V. L. Petrović and D. M. El Mezeni, ``Reduced-complexity offset Min-sum based layered decoding for 5G LDPC codes,'' {\it Proc. Telecomm. Forum}, pp. 1-4, Nov 2020.

\bibitem{ParaLayeredLDPCArch3}
K. K. Gunnam, G. S. Choi, and M. B. Yeary, ``A parallel VLSI architecture for layered decoding for array LDPC codes,'' {\it Proc. IEEE Intl. Conf. on VLSI Design}, pp. 738-743, Feb 2007.

\bibitem{Liva}
H. Bartz and G. Liva, ``On decoding schemes for the MDPC-McEliece cryptosystem,'' {\it{Proc. of Intl. ITG Conf. on Syst., Commun., and Coding}}, pp. 1-6, Mar. 2019.

\bibitem{BIKE}
T. B. Paiva and R. Terada, ``Faster constant-time decoder for MDPC codes and applications to BIKE KEM,'' {\it IACR Trans. on Cryptographic Hardw. and Embedded Syst.}, vol. 2022, no. 4, pp. 110-134, Aug. 2022.

\bibitem{Baldi}
P. Santini, M. Battaglioni, M. Baldi, and F. Chiaraluce, ``Analysis of the error correction capability of LDPC and MDPC codes under parallel bit-flipping decoding and application to cryptography,'' {\it{IEEE Trans. on Commun.}}, vol. 68, no. 8, pp. 4648-4660, Aug. 2020.

\bibitem{ErrorFloor}
S. Arpin, {\it et al.} ``A study of error floor behavior in QC-MDPC codes,'' {\it Proc. Intl. Conf. on Post-Quantum Cryptogr.}, pp. 89-103, Sep. 2022.


\bibitem{GuneysuDATE}
I. V. Maurich and T. Güneysu, ``Lightweight code-based cryptography: QC-MDPC McEliece encryption on reconfigurable devices,'' {\it{ Proc. IEEE Design, Autom. \& Test in Eur. Conf. \& Exhib.}}, pp. 1-6, Mar. 2014.

\bibitem{Guneysu}
I. V. Maurich, T. Oder, and T. Güneysu, ``Implementing QC-MDPC McEliece encryption,''  {\it{ACM Trans. on Embedded Comput. Syst.}}, vol. 14, no. 3, pp. 1-27, Apr. 2015.

\bibitem{Hu}
J. Hu and R. Cheung, ``Area-time efficient computation of Niederreiter encryption on QC-MDPC codes for embedded hardware,'' {\it{IEEE Trans. on Computers}}, vol. 66, no. 8, pp. 1313-1325, Aug. 2017.

\bibitem{XieSparse}
Z. Xie and X. Zhang, ``Sparsity-aware medium-density parity-check decoder for McEliece cryptosystems," {\it{IEEE Trans. on Circuits and Syst.-II}}, vol. 70, no. 9, pp. 3343-3347, Sep. 2023.

\bibitem{BaldiAttack}
P. Santini, M. Battaglioni, F. Chiaraluce, and M. Baldi, ``Analysis of reaction and timing attacks against cryptosystems based on sparse parity-check codes," {\it Proc. Code-Based Cryptogr.: 7th Intl. Workshop}. pp. 115-136, Jul. 2019.

\bibitem{Reaction}
Q. Guo, T. Johansson, and P. Stankovski, ``A key recovery attack on MDPC with CCA
security using decoding errors," {\it Proc. ASIACRYPT}, pp. 789-815, Dec. 2016.

\bibitem{Minsum} 
J. Chen and M. Fossorier, ``Density evolution for two improved BP-based decoding algorithms of LDPC codes," {\it
IEEE Commun. Lett.}, vol. 6, no. 5, pp. 208-210, May 2002.

\bibitem{McElieceSlicedMP}
J. Cai and X. Zhang, ``Low-Complexity parallel Min-sum medium-density parity-check decoder for McEliece cryptosystem," {\it{IEEE Trans. on Circuits and Syst.-I}}, Sep. 2023.

\bibitem{SlicedLDPC}
L. Liu and C.-J. Shi, ``Sliced message passing: high throughput overlapped
decoding of high-rate low-density parity-check Codes,'' {\it IEEE
Trans. on Circuits and Syst.-I}, vol. 55, no. 11, pp. 3697-3710, Dec. 2008.

\bibitem{LayeredLDPC}
M. Mansour and N. Shanbhag, ``A 640-Mb/s 2048-bit programmable LDPC decoder chip,'' {\it IEEE Journ. of Solid-State Circuits}, vol. 41, no. 3, pp. 684-698, Mar. 2006.

\bibitem{MyBook}
X. Zhang, {\it VLSI Architectures for Modern Error Correcting Codes}, CRC Press, Jul. 2015.

\bibitem{MultiBlockRow}
X. Zhang and Y. Tai, ``High-speed multi-block-row layered decoding for Quasi-cyclic LDPC codes,'' {\it Proc. IEEE Global Conf. on Signal and Info. Processing}, pp. 11-14, Dec. 2014.



\end{thebibliography}
\end{document}